\documentclass{INTERSPEECH2023}

\interspeechcameraready

\usepackage{multicol}
\usepackage{multirow}
\usepackage{color, colortbl}
\usepackage{xcolor}
\colorlet{eng}{blue!10}
\colorlet{cmn}{teal!10}
\colorlet{multi}{yellow!10}
\colorlet{euro}{orange!10}
\usepackage{pifont}
\usepackage{hyperref}
\usepackage{graphicx}
\usepackage{subcaption}
\usepackage{amsmath,amsfonts,bm}
\DeclareMathAlphabet{\mathsfit}{\encodingdefault}{\sfdefault}{m}{sl}
\SetMathAlphabet{\mathsfit}{bold}{\encodingdefault}{\sfdefault}{bx}{n}
\newcommand{\tens}[1]{\bm{\mathsfit{#1}}}
\usepackage{tablefootnote}

\title{MMM: Multi-Layer Multi-Residual Multi-Stream Discrete Speech Representation from Self-supervised Learning Model}
\name[affiliation={1}]{Jiatong}{Shi}
\name[affiliation={2}]{Xutai}{Ma}
\name[affiliation={2}]{Hirofumi}{Inaguma}
\name[affiliation={2}]{Anna}{Sun}
\name[affiliation={1}]{Shinji}{Watanabe}
\address{$^{1}$ Carnegie Mellon University, $^{2}$ Meta AI }
\email{\{jiatongs, swatanab\}@cs.cmu.edu}

\usepackage[
backend=biber,
style=ieee,
citestyle=numeric-comp,
maxbibnames=3,
maxcitenames=3,
doi=false,isbn=false,url=false,eprint=false
]{biblatex}

\defbibheading{bibliography}[\refname]{}

\DeclareSourcemap{
	\maps[datatype=bibtex, overwrite=true]{
		\map{
			\step[fieldsource=booktitle,
			match=\regexp{.*Interspeech.*},
			replace={Proc. Interspeech}]
			\step[fieldsource=journal,
			match=\regexp{.*INTERSPEECH.*},
			replace={Proc. Interspeech}]
			\step[fieldsource=booktitle,
			match=\regexp{.*ICASSP.*},
			replace={Proc. ICASSP}]
			\step[fieldsource=booktitle,
			match=\regexp{.*icassp_inpress.*},
			replace={Proc. ICASSP (in press)}]
			\step[fieldsource=booktitle,
			match=\regexp{.*Acoustics,.*Speech.*and.*Signal.*Processing.*},
			replace={Proc. ICASSP}]
			\step[fieldsource=booktitle,
			match=\regexp{.*International.*Conference.*on.*Learning.*Representations.*},
			replace={Proc. ICLR}]
			\step[fieldsource=booktitle,
			match=\regexp{.*International.*Conference.*on.*Computational.*Linguistics.*},
			replace={Proc. COLING}]
			\step[fieldsource=booktitle,
			match=\regexp{.*SIGdial.*Meeting.*on.*Discourse.*and.*Dialogue.*},
			replace={Proc. SIGDIAL}]
			\step[fieldsource=booktitle,
			match=\regexp{.*International.*Conference.*on.*Machine.*Learning.*},
			replace={Proc. ICML}]
			\step[fieldsource=booktitle,
			match=\regexp{.*North.*American.*Chapter.*of.*the.*Association.*for.*Computational.*Linguistics:.*Human.*Language.*Technologies.*},
			replace={Proc. NAACL}]
			\step[fieldsource=booktitle,
			match=\regexp{.*Empirical.*Methods.*in.*Natural.*Language.*Processing.*},
			replace={Proc. EMNLP}]
			\step[fieldsource=booktitle,
			match=\regexp{.*Association.*for.*Computational.*Linguistics.*},
			replace={Proc. ACL}]
			\step[fieldsource=booktitle,
			match=\regexp{.*Automatic.*Speech.*Recognition.*and.*Understanding.*},
			replace={Proc. ASRU}]
			\step[fieldsource=booktitle,
			match=\regexp{.*Spoken.*Language.*Technology.*},
			replace={Proc. SLT}]
			\step[fieldsource=booktitle,
			match=\regexp{.*Speech.*Synthesis.*Workshop.*},
			replace={Proc. SSW}]
			\step[fieldsource=booktitle,
			match=\regexp{.*workshop.*on.*speech.*synthesis.*},
			replace={Proc. SSW}]
			\step[fieldsource=booktitle,
			match=\regexp{.*Advances.*in.*neural.*information.*processing.*},
			replace={Proc. NeurIPS}]
			\step[fieldsource=booktitle,
			match=\regexp{.*Advances.*in.*Neural.*Information.*Processing.*},
			replace={Proc. NeurIPS}]
			\step[fieldsource=booktitle,
			match=\regexp{.*Workshop.*on.* Applications.* of.* Signal.*Processing.*to.*Audio.*and.*Acoustics.*},
			replace={Proc. WASPAA}]
			\step[fieldsource=publisher,
			match=\regexp{.+},
			replace={{}}]
			\step[fieldsource=month,
			match=\regexp{.+},
			replace={{}}]
			\step[fieldsource=location,
			match=\regexp{.+},
			replace={{}}]
			\step[fieldsource=address,
			match=\regexp{.+},
			replace={{}}]
			\step[fieldsource=organization,
			match=\regexp{.+},
			replace={{}}]
			\step[fieldsource=pages,
			match=\regexp{.+},
			replace={{}}]
		}
	}
}


\begin{document}

\maketitle
 
\begin{abstract}
Speech discrete representation has proven effective in various downstream applications due to its superior compression rate of the waveform, fast convergence during training, and compatibility with other modalities. Discrete units extracted from self-supervised learning (SSL) models have emerged as a prominent approach for obtaining speech discrete representation. However, while discrete units have shown effectiveness compared to spectral features, they still lag behind continuous SSL representations. In this work, we propose MMM, a multi-layer multi-residual multi-stream discrete units extraction method from SSL. Specifically, we introduce iterative residual vector quantization with K-means for different layers in an SSL model to extract multi-stream speech discrete representation. Through extensive experiments in speech recognition, speech resynthesis, and text-to-speech, we demonstrate the proposed MMM can surpass or on-par with neural codec's performance under various conditions.

\end{abstract}

\section{Introduction}
\label{sec: intro}

Efficient representation of speech signals is fundamental for a wide array of speech processing tasks, ranging from automatic speech recognition (ASR) to text-to-speech (TTS) synthesis. Traditionally, spectral features such as linear spectrograms or mel spectrograms have been extensively used in speech processing due to their robustness and interpretability \cite{hermansky1990perceptual, huang2001spoken}. However, with the advent of deep learning, there has been a paradigm shift towards utilizing neural networks as feature extractors, offering improved performance over traditional methods \cite{sainath15_interspeech, hermansky2000tandem}.

More recently, self-supervised learning (SSL) approaches have gained prominence in speech representation learning. These methods leverage large amounts of unlabeled speech data to learn powerful representations, surpassing previous state-of-the-art results on various benchmarks \cite{baevski2020wav2vec, hsu2021hubert, chen2022wavlm, babu22_interspeech, chiu2022self, shi23h_interspeech, shi2024multiresolution, superb, mlsuperb}. However, continuous SSL representations often suffer from scalability issues in terms of storage, computation, and integration with other modalities \cite{chang2022distilhubert, chang23b_interspeech, chang2023exploring, shi2023bridging}. This has led to a growing interest in discrete speech representation approaches, which offer more efficient and compact representations.

Two prominent methods that have emerged are SSL-based units and neural audio codecs. SSL-based units leverage clustering methods in an unsupervised manner to convert continuous SSL representations into discrete units, initially explored for speech resynthesis and subsequently proven effective in tasks such as speech translation (ST), ASR, TTS, and spoken language understanding \cite{hayashi2020discretalk,polyak21_interspeech, lee2022direct, chang2023exploring, zhang2023dub, shi2023enhancing, chang23b_interspeech, chang2023exploring, barrault2023seamless, yan-etal-2023-espnet, yang2023towards}. However, while SSL-based units offer efficiency and effectiveness benefits, they often fall short of achieving better performance than continuous SSL representations and lack detailed acoustics for speech generation purposes \cite{chang23b_interspeech, barrault2023seamless, nguyen2023expresso}.

On the other hand, neural audio codecs focus on audio resynthesis tasks, employing neural networks to learn auto-encoder architectures for discrete codec generation. A key component of recent neural codec methods is the use of residual vector quantization (RVQ) in the discretization process, resulting in multi-stream audio compression that retains subtle audio details with enhanced expressiveness \cite{zeghidour2021soundstream, defossez2022high, yang2023hifi, nguyen2023expresso}. This property has led to extensions of neural audio codecs to text-to-speech and spoken language models, demonstrating robust speech generation capabilities in zero-shot multi-speaker TTS scenarios \cite{wang2023neural, wang2023speechx, yang2023uniaudio}. However, neural codecs optimized for resynthesis tasks often lack semantic information due to their focus on streaming efficiency and short-context representation \cite{chang2023exploring}.

Despite their differences, limited comparative studies exist between SSL-based units and neural audio codecs under comparable conditions. Notably, SSL-based units typically operate in a single-stream setting, which offers less information capacity compared to multi-stream codecs.

This study propose a multi-layer multi-residual multi-stream (MMM) framework to extract discrete speech representation from continuous SSL representations. Specifically, we conduct RVQ-style quantization with K-means clustering to enable multi-stream discrete tokens in each single SSL layer. By combining with streams from multiple SSL layers, we further enhance the richness of the SSL-based units. With extensive experiments in ASR, we reveal that the proposed MMM-based discrete speech units elevates the performance of original single-stream SSL by a large margin and almost approaches the top-line performance with continuous SSL representation. While maintaining better ASR performance, we also demonstrate that the MMM-based units could achieve comparable or better performance to the neural codec-based approach in speech resynthesis and TTS.


\section{Methodology}
\label{sec: method}

SSL discrete units are derived from a single-layer hidden representation within a specific SSL model. Given a speech signal $\bm{x}$, we represent an SSL model as $\mathrm{S}$, which produces layer-wise representations denoted as $\tens{R} = [\bm{R}^1, \ldots, \bm{R}^L]$, where $L$ signifies the number of layers in $\mathrm{S}$. For a particular layer $l$, each element $\bm{r}^{l} \in \bm{R}^l$ comprises a sequence of vectors $[\bm{r}^{l}_1, \ldots, \bm{r}^{l}_T]$, with $T$ representing the frame count. Upon selecting step $t$ for analysis, we employ a $l$-th layer K-means model as $\mathrm{K}^l$ to determine $K^l$ cluster centroids. 
With these cluster centroids obtained by K-means training, we can cluster the feature vectors $\bm{r}_t^l$ by finding the optimal cluster index, which minimizes the Euclidean distance between $\bm{r}_t^l$ and each cluster centroid $\bm{c}_{k} ^l$ as:
\begin{equation}
    \label{eq:k-means}
    k_t ^l = \mathrm{argmin}_{k \in {1, \ldots, K^l}} ||\bm{r}_t ^l - \bm{c}_{k} ^l||^2.
\end{equation}
The final cluster IDs $[k_1^l, \ldots, k_T^l]$ serve as discrete units at particularly layer $l$ for subsequent downstream tasks

In this study, we broaden the application from a single-stream scenario to encompass multi-stream scenarios. To achieve this, we propose two complementary strategies for multi-stream modeling, which involve leveraging either a single layer or multiple layers from the SSL model $\mathrm{S}$.

\subsection{Multi-stream from a single layer}
\label{ssec: rvq}


The first approach concentrates on generating multiple streams from a single layer within the SSL model.
Citing \cite{shi23h_interspeech}, it's demonstrated that the single-layer continuous representation, specifically from a HuBERT-base model, contains sufficient detail for audio resynthesis at frame resolutions of both 40ms and 100ms. Nonetheless, the resynthesis audio quality markedly deteriorates following quantization via K-means. This decline highlights a significant limitation: discrete units derived from K-means quantization tend to omit intricate acoustic details originally present in the waveform signals.

To counteract the information loss inherent in quantization, our method involves estimating additional streams of discrete units. This process is aligned with the principle of RVQ, adhering to the unsupervised nature of the original K-means-based approach. 
For training, in the $l$-th layer, we iteratively estimate $m$-th K-means model $\mathrm{K}^{m, l}$ on the residual feature from previous K-means models $[\mathrm{K}^{1, l}, \ldots, \mathrm{K}^{m-1, l}]$. During inference, the cluster index $k_{t}^{m, l}$ in frame $t$ can then be obtained through iterative procedures. The following equations elaborate the detailed inference steps from the first stream to stream $m$.

\begin{align}
    k_{t}^{1, l} & = \mathrm{argmin}_{k \in {1,,..., K^{1, l}}} \left|\left| \bm{r}_t ^l  -  \bm{c}_{k}^{1, l} \right|\right|^2, \\
    k_{j}^{2, l} & = \mathrm{argmin}_{k \in {1,,..., K^{2, l}}} \left|\left| (\bm{r}^l_t - \bm{c}_{t}^{1, l}) - \bm{c}_{k}^{2, l} \right|\right|^2, \\
    & \vdots \notag \\
        k_{j}^{m, l} & = \mathrm{argmin}_{k \in {1,,..., K^{m, l}}} \left|\left| \left(\bm{r}^l_t - \sum_{u=1}^{m-1} \bm{c}_{t}^{u, l} \right) - \bm{c}_{k}^{m, l} \right|\right|^2,
\end{align}
where we define $\bm{c}_{t}^{m, l}$ as the selected centroid at $t$, i.e., $\bm{c}_{t}^{m, l} := \bm{c}_{k}^{m, l}|_{k = k_{t}^{m, l}}$
.
Estimated $k_t^{m,l}$ becomes the discrete unit of the $m$-th stream at $t$-th frame.


\subsection{Multi-stream from multiple layers}
\label{ssec: multi-layer}


While Section~\ref{ssec: rvq} elaborated on enhancing discrete representations through a single SSL model layer $l$, an alternative strategy employs multi-layer representations. Historically, the integration of information across multiple layers in SSL models has proven invaluable, particularly when leveraging frozen SSL representations for downstream tasks. This is exemplified in the Speech Universal PERformance Benchmark (SUPERB), where layer-wise representations are combined through a weighted sum approach, achieving commendable results across a variety of speech processing tasks \cite{superb}. Subsequent research further validates this approach, demonstrating that different layers of an SSL model encapsulate distinct facets of speech-related information \cite{pasad2021layer, chang2021exploration, chen2022wavlm, mlsuperb, shi23h_interspeech, shi2024multiresolution}.

In the context of discrete representations, the significance of utilizing multi-layer information becomes even more pronounced. Reliance on a single layer for information extraction may inadvertently prioritize certain features while overlooking others, a discrepancy that becomes especially noticeable following the quantization process (see Section~\ref{ssec: rvq}). Therefore, a multi-layered approach not only diversifies the extracted information but also mitigates the risk of information bias, ensuring a more holistic representation of speech signals.

The formulation of multi-layer multi-stream discrete representations can be easily extracted from continuous representation $\tens{R}$. 
To be specific, we specify $L'$ layers from the multi-layer representation $\tens{R}$. Then, for the K-means model $\mathrm{K}^{m, l'}$ of a selected layer $l'$, the designated cluster IDs $[k^{m, l'}_1, \ldots, k^{m, l'}_T]$ can be used as the discrete token for downstream tasks. The final MMM-based representation with $M \times L'$ streams can be obtained by combining the multi-stream extraction methods in Section~\ref{ssec: rvq} and Section~\ref{ssec: multi-layer}.

\subsection{Application of MMM-based discrete units}
\label{ssec: application}

Building on existing literature using discrete representations~\cite{polyak21_interspeech, lee2022direct, wang2023neural, chang2023exploring, chang23b_interspeech, borsos2023soundstorm, wang2023speechx, yang2023uniaudio, wu2024codec, yang2023towards}, we explore the application of MMM-based discrete units as either inputs or outputs across various applications. Our investigation encompasses both scenarios, employing a strategic approach to integrate discrete representations into downstream tasks.

\textbf{Input Scenario}: For representations used as inputs, we initially transform these discrete entities into embeddings. Subsequently, we apply a summation across different streams to integrate these embeddings. Specifically, for streams derived from a single layer, a direct summation of embeddings is executed, mirroring the inverse operation of the RVQ technique. Conversely, when dealing with streams from multiple layers, we adopt a learnable weighted summation for their aggregation. This method leverages learnable weights, optimized through a $\mathrm{Softmax}$ function, aligning with strategies from the SUPERB series~\cite{superb, mlsuperb, feng2023superb, tsai2022superb, shi2023findings}. This approach ensures that the integration of multi-layer streams is both dynamic and informed by the data. In cases where streams are sourced from both RVQ and multiple layers, we first combine embeddings from identical layers before proceeding to merge across different layers.

\textbf{Output Scenario}: For outputs utilizing MMM-based units, we implement a straightforward parallel approach that independently predicts various streams of tokens. This decision is different from prior work on neural audio codecs, where it was noted that models trained on RVQ necessitate auto-regressive modeling to maintain decoder quality~\cite{wang2023neural, copet2024simple, yang2023uniaudio}. Despite this, our pilot experiments indicate that the impact of such modeling on SSL-based units is minimal, thereby justifying our preference for a simpler, parallel prediction method for MMM-based discrete tokens.

\section{Experiments}

As discussed in Section~\ref{ssec: application}, the discrete token can be applied in both input and output scenarios. For the input scenario, we select the two classical tasks of the proposed method: ASR and speech resynthesis (i.e., vocoder). The two tasks consider both usages in understanding and generation. For the output scenario, we conduct TTS, which serves as the backbone of most other tasks that produce discrete units~\cite{barrault2023seamless, maiti2023voxtlm}.

\subsection{Speech recognition}
\label{sec:asr_ex}

\noindent \textbf{Dataset}: Our ASR experiments align with the discrete speech challenge at Interspeech2024.\footnote{\scriptsize{\url{https://www.wavlab.org/activities/2024/Interspeech2024-Discrete-Speech-Unit-Challenge/}}} We utilize a dataset comprising Librispeech's train-clean-100h \cite{panayotov2015librispeech} combined with the ML-SUPERB multilingual 1-hour set \cite{mlsuperb}. This blend enables the examination of both clean English read speech and multilingual ASR tasks. The total dataset spans 310.4 hours and encompasses 143 languages, offering a broad spectrum for evaluation.

\noindent \textbf{SSL Models}: Informed by insights from \cite{chang23b_interspeech, mlsuperb, shi2023findings} and the challenge's baseline, we recognize the distinct performance capabilities of WavLM~\cite{chen2022wavlm} and XLS-R~\cite{babu22_interspeech} across different corpora. WavLM-large exhibits notable effectiveness in English datasets, whereas XLS-R (300M version) is better suited for multilingual datasets. To comprehensively assess performance across both English and multilingual corpora, we employ these models as our primary SSL candidates.

\noindent \textbf{Clustering and Downstream Settings}: Consistent with the challenge's baseline, we opt for a random subsampling of 30\% of the training set utterances for K-means clustering, setting the cluster size at 500 for each stream, i.e., $K^{m, l}=500$. Deviating from the baseline, our methodology eschews additional byte-pair encoding and deduplication, simplifying the alignment of different streams. The downstream model leverages the same encoder-decoder architecture within ESPnet~\cite{watanabe2018espnet} as the challenge baseline. 

\noindent \textbf{Proposed Method}: 
Following the methodology outlined in Section~\ref{sec: method}, we extract multi-stream tokens from both a single layer and multiple layers. For the single-layer approach, two ($M=2$) streams of discrete representations are extracted from the 21st layer of both WavLM and XLS-R. For multi-layer scenarios, we select four ($L'=4$) layers (\{9, 15, 21, 22\}) to balance compression efficiency and performance. As in Section~\ref{ssec: multi-layer}, the two methods can be combined to yield eight streams of tokens.

\noindent \textbf{Baseline and Ablation Studies}: Our baseline comparison uses the single-layer, one-stream SSL units as defined in the challenge. Additionally, we use the publicly-availble Encodec-24kHz model~\cite{defossez2022high} with 8 streams as another baseline. Ablation studies are conducted for both single-layer (with $M={2, 4, 8}$ streams) and multi-layer configurations, including a variety of layer selections. Layer selection for multi-layer scenarios is optimized using a model with all layers, employing learnable weighted summation (see Section~\ref{ssec: application}), and then selecting the top four weighted layers for further analysis. 

\begin{table}[!t]
    \centering
    \caption{ASR performance on discrete speech challenge dataset. ``$+$" indicates single-layer multi-stream setting and ``$\dagger$" stands for multi-layer multi-stream setting. $M/L'$ corresponding to number layers in single/multi-layer multi-stream settings.}
    \vspace{-5pt}
    \resizebox{\linewidth}{!}{
    \begin{tabular}{l|c|cc|c} 
        \toprule
        \multirow{2}{*}{SSL} & Streams & 
         Librispeech & ML-SUPERB (1h) & Bitrate \\ 
         & $M \times L'$ & WER & CER \\
        \midrule
         WavLM & 1 * 1 & 6.3 & 22.8 & 548.3 \\
         XLS-R & 1 * 1 & 14.1 & 21.4 & 548.3 \\
         Encodec & 8 & 15.9 & 35.9 & 6000.0 \\
         \midrule
         WavLM$+$ & 2 * 1 & 5.9 & 21.4 & 1096.6 \\
         WavLM$\dagger$ & 1 * 4 & 5.0 & 20.8 & 2193.2 \\
         XLS-R$+$ & 2 * 1 & 10.5 & 19.3 & 1096.6 \\
         XLS-R$\dagger$ & 1 * 4 & 7.3 & 18.0 & 2193.2 \\
         \midrule
         WavLM$+\dagger$ & 2 * 4 & \textbf{4.7} & 19.5 & 4386.3 \\
         XLS-R$+\dagger$ & 2 * 4 & 6.8 & \textbf{17.5} & 4386.3 \\
        \bottomrule
    \end{tabular}
    }
    \label{tab:asr}
    \vspace{-10pt}
\end{table}

\begin{table}[!ht]
    \centering
    \caption{ASR ablation studies on the single-layer setting over discrete speech challenge dataset.}
    \vspace{-5pt}
    \resizebox{\linewidth}{!}{
    \begin{tabular}{l|c|cc|c} 
        \toprule
        \multirow{2}{*}{SSL} & Streams & 
         Librispeech & ML-SUPERB (1h) & \multirow{2}{*}{Bitrate} \\ 
         & $M$ & WER & CER \\
        \midrule
         WavLM$+$ & 1 & 6.3 & 22.8 & 548.3 \\
         WavLM$+$ & 4 & 6.1 & 21.5 & 2193.2 \\
         WavLM$+$ & 8 & 6.4 & 21.7 & 4386.3 \\
         \midrule
         WavLM$+$ & 2 & \textbf{5.9} & \textbf{21.4} & 1096.6 \\
        \bottomrule
    \end{tabular}
    }
    \label{tab:asr-2}
    \vspace{-10pt}
\end{table}

\begin{table}[!ht]
    \centering
    \caption{ASR ablation studies on the multi-layer setting over discrete speech challenge dataset. Detailed layer indexes are shown in the ``Layers" column.}
    \vspace{-5pt}
    \resizebox{\linewidth}{!}{
    \begin{tabular}{l|c|cc|c} 
        \toprule
        \multirow{2}{*}{SSL} & Layers & 
         Librispeech & ML-SUPERB (1h) & \multirow{2}{*}{Bitrate} \\ 
         & $L'$ & WER & CER \\
        \midrule
         WavLM$\dagger$ & 21 & 6.3 & 22.8 & 548.3 \\
         WavLM$\dagger$ & 1-4 & 6.8 & 27.7 & 2193.2  \\
         WavLM$\dagger$ & 11-14 & 6.1 & 21.9 & 2193.2  \\
         WavLM$\dagger$& 21-24 & 5.5 & 21.5 & 2193.2  \\
         \midrule
         WavLM$\dagger$ & 0-24 & \textbf{4.9} & \textbf{19.9} & 13707.2 \\
         \midrule
         WavLM$\dagger$ & 9, 15, 21, 22 & 5.0 & 20.8 & 2193.2 \\
        \bottomrule
    \end{tabular}
    }
    \label{tab:asr-3}
    \vspace{-10pt}
\end{table}

\noindent \textbf{Evaluation Metrics}: We follow the setting in the discrete challenge for evaluation metrics, including average word error rate~(WER) for Librispeech test sets and weighted average character error rate (CER) for ML-SUPERB test sets (i.e., normal test set and few-shot test set). Bitrate is also reported, following the discrete speech challenge guidelines.

\noindent \textbf{Results and Discussion}: The main results, as depicted in Table~\ref{tab:asr}, illustrate that our proposed method uniformly enhances performance across both Librispeech and ML-SUPERB datasets for both SSL models. Additionally, the performance improvements yielded by the two proposed approaches appear to be complementary. Compared to Encodec tokens, even the single-stream discrete token sequences have better performances. The finding is aligned with the observations in \cite{chang2023exploring} where SSL-based discrete representations outperform the neural codec-based method, specifically Encodec.

Our detailed ablation studies focusing on the number of streams are summarized in Tables~\ref{tab:asr-2} and~\ref{tab:asr-3}. In the context of the single-layer multi-stream approach, augmenting the number of streams does not always lead to performance enhancement. As highlighted in Table~\ref{tab:asr-2}, the ASR performance drops when the stream count is increased to 4 or 8, compared to the two-stream scenario. This deterioration could be attributed to the potential instability associated with employing K-means for higher-order residual information extraction. Conversely, for the multi-layer multi-stream scenario, optimal performance is attained when all layers are utilized at the cost of a higher bitrate. 
By selecting the layers with the top four weights, we show that it's possible to maintain performance levels with only a slight degradation while achieving a significant reduction in the bitrate.

\subsection{Speech resynthesis (vocoder)}
\label{ssec: vocoder}

\begin{table*}[ht]
    \centering
    \caption{Speech resynthesis performance on discrete speech challenge dataset (filtered Expresso). ``$+$" indicates single-layer multi-stream setting and ``$\dagger$" stands for multi-layer multi-stream setting. $M/L'$ corresponding to number layers in single/multi-layer multi-stream settings.}
    \vspace{-5pt}
    \resizebox{0.55\linewidth}{!}{
    \begin{tabular}{l|c|ccc|c} 
        \toprule
        SSL & Streams ($M \times L'$) & MCD & F0 RMSE & UTMOS & Bitrate \\ 
        \midrule
         HuBERT  & 1 * 1 & 7.19 & 0.42 & 2.27 & 448.3 \\
         Encodec & 8 & \textbf{3.91} & 0.21 & 3.18 & 3586.4 \\
         \midrule
         HuBERT (\textit{S}) & 2 * 1 & 6.79 & 0.32 & 2.89 & 896.6 \\
         HuBERT (\textit{M}) & 1 * 4 & 5.12 & 0.22 & 3.10 & 1793.2 \\
         \midrule
         HuBERT (\textit{S+M}) & 2 * 4 & 4.54 & \textbf{0.20} & \textbf{3.22} & 3586.4 \\
        \bottomrule
    \end{tabular}
    }
    \label{tab:vocoder}
\end{table*}
\begin{table*}[ht]
    \centering
    \caption{TTS performance on discrete speech challenge dataset (LJSpeech). }
    \vspace{-5pt}
    \resizebox{0.6\linewidth}{!}{
    \begin{tabular}{l|c|cccc|c} 
        \toprule
        SSL & Streams ($M \times L'$)  & MCD & F0 RMSE & WER & UTMOS & Bitrate \\ 
        \midrule
         HuBERT  & 1 * 1 &   7.19 & 0.26 & 8.1 & 3.73 & 448.3 \\
         Encodec & 8 & \textbf{7.01} & 0.29 & 7.8 & 4.01 & 3586.4 \\
         \midrule
         HuBERT (\textit{S}) & 2 * 1 & 7.11 & 0.29 & 8.0 & 3.79 & 896.6 \\
         HuBERT (\textit{M}) & 1 * 4 & 7.25 & \textbf{0.24} & \textbf{7.7} & 4.06 & 1793.2 \\
         \midrule
         HuBERT (\textit{S+M}) & 2 * 4 & 7.15 & 0.25 & \textbf{7.7} & \textbf{4.15} & 3586.4  \\
        \bottomrule
    \end{tabular}
    }
    \label{tab:tts}
    \vspace{-10pt}
\end{table*}


\noindent \textbf{Dataset}: 
Our speech resynthesis experiments are anchored in the TTS (vocoder) track of the discrete speech challenge at Interspeech 2024. We utilize a curated subset of the Expresso benchmark~\cite{nguyen2023expresso}, focusing on a single-speaker dataset and excluding segments with singing, overlapping speech, and long-form content. The experiments adhere to the official train-dev-test partitioning provided by the challenge organizers.\footnote{\scriptsize{\url{https://github.com/ftshijt/Interspeech2024_DiscreteSpeechChallenge}}} To align with the original SSL model's specifications, audio samples at 48kHz are downsampled to 16kHz for compatibility.


\noindent \textbf{Experimental Set-ups}: 
Echoing prior studies~\cite{polyak21_interspeech, lee2022direct, yan-etal-2023-espnet, maiti2023voxtlm, nguyen2023expresso} in discrete-based speech resynthesis that predominantly utilize HuBERT~\cite{hsu2021hubert}, our experiments also employ a pre-trained HuBERT-base model trained on the full Librispeech dataset via S3PRL\cite{superb}. Clustering is performed on the respective training sets with a designated cluster size of 500 per stream. The discrete HiFiGAN model serves as our downstream backbone, configured in accordance with \cite{yan-etal-2023-espnet}.
For our baseline, the 9th layer is selected for extracting single-stream SSL units, drawing from the methodology in \cite{polyak21_interspeech, lee2022direct, yan-etal-2023-espnet}. In exploring multi-stream capabilities, we investigate two ($M=2$) streams for single-layer scenarios, four ($L'=4$) streams for multi-layer configurations\footnote{Layer 6, 9, 11, 12 are used, following the same selection strategy as the ASR track. We use the same setting for TTS experiments.}, and an integrated approach yielding eight streams ($M \times L' = 8$). 
We referred to these multi-stream configurations from the experimental findings in Section~\ref{sec:asr_ex}.
Additional Encodec baselines are trained on the same training set based on AudioCraft~\cite{copet2024simple}. We set the Encodec model with 8 streams at a 50Hz frame rate. For each stream, the codebook size is set to be 500, to be aligned with our experiments.


\noindent \textbf{Evaluation Metrics}: The evaluation framework prioritizes objective metrics in line with the discrete speech challenge's guideline. These metrics include mel cepstral distortion~(MCD), F0 root mean square error (F0 RMSE), UTMOS~\cite{saeki2022utmos}, and bitrate. The MCD and F0 RMSE metrics are calculated using the ESPnet-TTS toolkit~\cite{hayashi2020espnet, hayashi2021espnet2}.

\noindent \textbf{Results and Discussion}: Table~\ref{tab:vocoder} illustrates that our proposed method outperforms the baselines in UTMOS and F0 RMSE. Compared to the single-stream baseline, both of our proposed approaches not only improve all evaluated metrics, but also exhibit complementary benefits to each other. 
Notably, even when compared to Encodec, our method demonstrates superior UTMOS and F0 RMSE scores. This is particularly significant given that our discrete tokens are extracted unsupervisedly and are not explicitly optimized for the resynthesis task.

\subsection{Text-to-speech}

\noindent \textbf{Dataset}: For TTS, our examination leverages the LJSpeech dataset, a single-speaker female TTS corpus, in alignment with the discrete speech challenge recommendations. We rigorously adhere to the official dataset partitioning for the training set, as detailed by the challenge guidelines.

\noindent \textbf{Experimental Set-ups}: 
The TTS modeling experiments inherit several configurations from the speech resynthesis task. This includes utilizing the HuBERT-base SSL model, adopting a cluster size of 500 for K-means clustering, employing the entire training set for clustering, and integrating the discrete HiFi-GAN model. Distinctively, our downstream TTS model employs a modified VITS architecture, substituting its adversarial decoder with a transformer network designed to directly predict discrete units. This modification is informed based on the ESPnet VITS LJSpeech recipe and draws inspiration from \cite{hayashi2020discretalk}.\footnote{Hyperparameters are in line with \href{https://github.com/espnet/espnet/blob/tts2/egs2/ljspeech/tts1/conf/tuning/train_vits.yaml}{ESPnet VITS LJSpeech recipe}.}


\noindent \textbf{Evaluation Metrics}: The evaluation framework utilize MCD, F0 RMSE, UTMOS and bitrate. In addition, we also report CER as assessed by a pre-trained Whisper-large-V2 model~\cite{radford2023robust}.

\noindent \textbf{Results and Discussion}: The outcomes of the TTS experiments, as displayed in Table~\ref{tab:tts}, echo the observations made in the vocoder experiments detailed in Section~\ref{ssec: vocoder}. Compared to the Encodec model, we observe significant enhanced naturalness (reflected in improved UTMOS scores) with the TTS system trained on our proposed multi-stream tokens. This enhancement could be attributed to the SSL tokens possessing a richer semantic content, offering extra advantages in acoustic modeling when used as output. This semantic richness in SSL tokens likely facilitates a more effective acoustic representation for acoustic modeling, thereby enhancing the overall naturalness of the synthesized speech.

\section{Conclusion}

In this study, we reexamine the extraction of SSL-based discrete tokens, focusing on multi-stream modeling. We focus on two approaches: single-layer and multi-layer modeling. With extensive experiments across ASR, speech resynthesis, and TTS, we showcase that multi-stream tokens from SSL models can usually improve upon the single-stream SSL tokens. Moreover, we also find the proposed representations attain performance levels that are either superior to or on par with those achieved by neural codec methods. 

\section{Acknowledgements}
This work was supported by a Meta AI SRA grant. Jiatong Shi and Shinji Watanabe are funded in part of the Bridges2 system at PSC and Delta system at NCSA through allocations CIS210014 and IRI120008P from the ACCESS program, supported by NSF grants \#2138259, \#2138286, \#2138307, \#2137603, and \#2138296.



\section{References}
{
\printbibliography
}

\end{document}